\documentclass[aps,prc,twocolumn,showpacs,superscriptaddress,footinbib]{revtex4-1}

\usepackage{graphicx}
\usepackage{calrsfs}
\usepackage{dcolumn}
\usepackage{bm}
\usepackage{enumerate}
\usepackage{amsmath}

\usepackage[normalem]{ulem}  
\usepackage[colorlinks]{hyperref}

\begin{document}


\title{Isobar correlations bearing information on the properties of disassembling hot nuclear sources}

\author{S.R. Souza}
\affiliation{Instituto de F\'\i sica, Universidade Federal do Rio de Janeiro Cidade Universit\'aria, 
\\Caixa Postal 68528, 21941-972 Rio de Janeiro-RJ, Brazil}
\affiliation{Departamento de F\'\i sica, ICEx, Universidade Federal de Minas Gerais,
\\Av.\ Ant\^onio Carlos, 6627, 31270-901 Belo Horizonte-MG, Brazil}
\author{R. Donangelo}
\affiliation{Instituto de F\'\i sica, Universidade Federal do Rio de Janeiro Cidade Universit\'aria, 
\\Caixa Postal 68528, 21941-972 Rio de Janeiro-RJ, Brazil}
\affiliation{Instituto de F\'\i sica, Facultad de Ingenier\'\i a, Universidad de la Rep\'ublica, 
Julio Herrera y Reissig 565, 11.300 Montevideo, Uruguay}

\date{\today}

\begin{abstract}
Two-particle correlations based on the multiplicity of selected isobars are found to be sensitive to the parameterization of the fragments' binding energies and the breakup volume assumed in the model calculations.
The properties of these correlations have been examined in the framework of the Statistical Multifragmentation Model as a function of the breakup temperature.
The model calculations suggest that the maximum of these correlation functions occurs at well separated temperatures as the breakup volumes used in the model varies from 3 to 6 times that at normal density.
This is within the range assumed in most statistical calculations and supported by experiments.
Besides the position, the height and width of the maximum is also found to be sensitive to the parameterization of the fragments' binding energy.
The magnitude of the effects also depends on the selected isobars.
We found that, due to an interplay between the symmetry energy and the volume dependent terms of the Helmholtz free energy, ratios involving light mirror nuclei seem to enhance the effects in the case of nearly symmetric sources.
We suggest that the proposed correlation functions be used to obtain information on the fragments'  energy and on the breakup volume of nuclear sources.

\end{abstract}

\maketitle

\begin{section}{Introduction}
\label{sect:introduction}
Nuclear matter at the extreme conditions found in different stages of stelar evolution \cite{Bethe1990,LattimerDenseMatter1985,supernovae2009,BotvinaSNova2004,BotvinaSNova2005} may be recreated in experiments envolving central and mid-central collisions between heavy-ions at energies well above the Coulomb barrier \cite{BorderiePhaseTransition2008,reviewSubal2001,BonaseraBUU,exoticDens,PhaseTransitionBorderie2019}.
Such reactions provide, in this way, a means to study the properties of nuclear matter far from the equilibrium configuration, {\it i.e.}, its Equation of State (EOS).
In this context, many studies have been carried out over the last decades \cite{BotvinaMishustin2006,BorderiePhaseTransition2008,PhaseTransitionBorderie2019,reviewSubal2001,AMDReview2004} and important efforts have been made in order to determine such properties, as the symmetry energy EOS \cite{PhaseTransitionBorderie2019,SymEnergyBaoAnLi2006,isoNatowitz2007_2,symEnergyShatty2007,EOSLi,isoSMMTF,SouliotisCsym2009,TanIsoEOS,TanisoDiff,smmtf1,isotemp} and the nuclear caloric curve 
\cite{ccGSI,ccMoretto1996,ccNatowitzHarm,ccMa2005,ccMa1997,ccAuger,ccDagostino,ccSharenberg,isocc,ccSerfling,ccXi}, for instance.

The scenario in which matter quickly expands, after attaining a compressed configuration in the initial stages of the collision, is supported by different models and experiments \cite{BotvinaMishustin2006,AMDReview2004,BonaseraBUU,XLargeSystems,SymEnergyBaoAnLi2006,exoticDens,bkDensViola,PhaseTransitionBorderie2019,freezeOut,freezeOut2,freezeOutPiantelli2005PLB,isoNatowitz2007_2}.
In one of the possible pictures, the system reaches a freeze out configuration, at which thermal and chemical equilibrium are attained and the multifragment production takes place \cite{PhaseTransitionBorderie2019,freezeOut,freezeOut2,freezeOutPiantelli2005PLB,isoNatowitz2007_2}.
A continuous fragment production, during a short time span in the expansion phase, is suggested by other dynamical calculations \cite{AMDReview2004,AMDfragmentEmission2006} as an alternative  view.
These two frameworks have also been merged into a hybrid treatment \cite{postBreakup2019,EnergySpectra2018} in which excited fragments are created in a prompt breakup within a breakup volume and slowly deexcite as they travel away from each other due to their initial thermal and (possibly) flow velocities, besides the Coulomb repulsion among them.

The volume occupied by the system at the freeze out configuration, {\it i.e.} the breakup volume, is an important input information to some of the calculations mentioned above, besides the temperature and source's isotopic composition.
Furthermore, its determination is key to the study of the nuclear EOS.
Therefore, experimental efforts to determine this quantity have been devoted by different groups \cite{PhaseTransitionBorderie2019,freezeOut,freezeOut2,freezeOutPiantelli2005PLB,isoNatowitz2007_2,reviewInterferometryKonrad1990,interferometryBetty1987,interferometryGiuseppe2002,interferometryGiuseppe2003}.
Such studies suggest that the freeze out is attained at densities ranging from 1/3 to 1/10 of the normal nuclear matter value.
This is compatible with the assumptions made in different statistical models \cite{Bondorf1995,smmIsobaric} and also found in dynamical calculations \cite{XLargeSystems}.

In this work we propose an observable which is fairly sensitive to the breakup volume assumed in the statistical calculations: correlations based on the multiplicities of light isobars.
In the framework of the Statistical Multifragmentation Model (SMM) \cite{smm1,smm2,smm4}, we examine the behaviour of this quantity as a function of the breakup temperature, assuming different isotopic compositions for the disassembling source.
We suggest that this observable may help to narrow the uncertainty on the breakup density.
These correlations also turn out to be sensitive to the assumptions made to the fragments' binding energy.
In this way, they may, therefore,  provide information on the latter in the freeze out configuration.

The remainder of this manuscript is organized as follows: a brief description of the model is made in Sect.\ \ref{sect:model}, where we derive the expressions used in Sect.\ \ref{sect:results}, which is devoted to the presentation and discussion of the main results.
Concluding remarks are drawn in Sect.\ \ref{sect:conclusions}.

\end{section}
 
\begin{section}{Theoretical framework}
\label{sect:model}
The SMM \cite {smm1,smm2,smm4} assumes that a source of mass and atomic numbers $A_0$ and $Z_0$, respectvely, is formed in a breakup volume $V$.
This source undergoes a prompt decay, producing fragments whose abundances are dictated by their corresponding statistical weights.
Mass and charge are strictly conserved.
We adopt the canonical version of the model, so that the source is also assumed to be formed at temperature $T$.
In this way, the statistical weight associated with a source $(A_0,Z_0)$ is given by:

\begin{equation}
\Omega_{A_0,Z_0}=\sum_{f\in F_0}\prod_{i\in f}\frac{\omega_i^{n_i}}{n_i!}\;,
\label{eq:Omega}
\end{equation}

\noindent
where $n_i$ symbolizes the multiplicity of the species `$i$' with mass and atomic numbers $a_i$ and $z_i$, respectively, and $f$ corresponds to a set of species so that

\begin{equation}
\sum_{i\in f} n_i z_i = Z_0 \;\; {\rm and}\;\; \sum_{i\in f} n_i a_i = A_0\;.
\label{eq:chargeMassCons}
\end{equation}

In the previous expression, $F_0$ represents the set of all partitions $f$ consistent with the constraint given by Eq.\ (\ref{eq:chargeMassCons}) and

\begin{equation}
\omega_i=\left(\frac{g_i V_f}{\lambda_T^3}A_i^{3/2}\right)
\exp\left(-{\cal F}_i/T\right)\;.
\label{eq:omega}
\end{equation}

\noindent
where $V_f=V-V_0$ is the free volume, $V_0$ is that at normal nuclear density, $\lambda_T=\sqrt{2\pi\hbar^2/mT}$, and $m$ is the nucleon mass.
The Helmholtz free energy ${\cal F}_i$, associated with the species,  has contributions from the fragment internal energy, besides those associated with the  Wigner-Seitz corrections \cite{smm1} to the Coulomb energy.
One should note that the term corresponding to the homogenous charged sphere of the Wigner-Seitz approximation does not play a role in the canonical formulation of the model, being a phase that cancels out in the relevant expressions and is, therefore, omitted.
Then, ${\cal F}_i$ reads:

\begin{equation}
{\cal F}_i={\cal F}^*_i-B_i-a_{\rm Coul} \frac{z^2_i}{a_i^{1/3}}\left(\frac{V_0}{V}\right)^{1/3}
\label{eq:fi}
\end{equation}

\noindent
where $a_{\rm Coul}$ is a model parameter associated with the Coulomb energy of the fragment, $B_i$ denotes its binding energy and ${\cal F}^*_i$ represents its internal Helmholtz free energy.
The parameters entering in Eq.\ (\ref{eq:fi}) are given in Refs.\ \cite{ISMMmass,ISMMlong} and we adopted the standard values for the parameters of ${\cal F}^*_i$, listed in \cite{ISMMlong}.
As two different parameterizations for $B_i$ are used in this work, we state in the next section the values adopted in each case.

The traditional SMM implementation employs a Monte Carlo sampling of the relevant partitions in order to calculate different observables.
In Refs.\ \cite{ChaseMekjian1995,SubalMekjian}, Das Gupta and Mekjian derived recurrence relations which allow the evaluation of the statistical weights very efficiently:

\begin{equation}
\Omega_{A_0,Z_0}=\sum_{i \in f_0}\frac{a_i}{A_0} \omega_i\,
\Omega_{A_0-a_i,Z_0-z_i}\;.
\label{eq:OmegaRec}
\end{equation}

\noindent
and $f_0$ represents the set of all possible species which may be produced.
From this expression, many obervables may be calculated exactly, in the framework of the model, such as the average species multiplicities, for instance.

Let us consider events in which fragments of species `$i$' and `$j$'  appear only once within each partition.
From Eq.\ (\ref{eq:Omega}), the probability of observing such partitions is:

\begin{equation}
Y_{i,j}=\frac{1}{\Omega_{A_0,Z_0}}\sum_{f\in F_0}{\omega_i}\delta_{n_i,1}\,{\omega_j}\delta_{n_j,1}\prod_{\substack{k\in f\\ k\ne i,j}}\frac{\omega_k^{n_k}}{n_k!}
\label{eq:yij0}
\end{equation}

\noindent
which may be rewritten as:

\begin{equation}
Y_{i,j}=\omega_i\omega_j \frac{\Omega^{(i,j)}_{A_0-a_i-a_j,Z_0-z_i-z_j}}{\Omega_{A_0,Z_0}}\;,
\label{eq:yij}
\end{equation}

\noindent
where

\begin{equation}
\Omega^{(i,j)}_{A_0-a,Z_0-z}\equiv\sum_{f\in F_{i,j}}\prod_{\substack{k\in f\\ k\ne i,j}}\frac{\omega_k^{n_k}}{n_k!}
\label{eq:omegaij}
\end{equation}

\noindent
and $F_{i,j}$ represents the set of partitions which fulfill the constraint expressed by Eq.\ (\ref{eq:chargeMassCons}), for $A_0\rightarrow A_0-a$ and $Z_0\rightarrow Z_0-z$, subject to the further constraint that fragments of species `$i$' and `$j$' appear only once.
The weight $\Omega^{(i,j)}_{A_0-a,Z_0-z}$ may be calculated using Eq.\ (\ref{eq:OmegaRec}) if one suppresses the species `$i$' and `$j$' in that expression.

In the same vein, the probability of observing partitions which have a single fragment of species `$i$' (`$j$') and none of species `$j$' (`$i$') is given by:

\begin{equation}
Y_\alpha = \sum_{f\in F_0}\frac{{\omega_\alpha}\delta_{n_\alpha,1}}{\Omega_{A_0,Z_0}}\prod_{\substack{k\in f\\ k\ne i,j}}\frac{\omega_k^{n_k}}{n_k!}\\
= \omega_\alpha\frac{\Omega^{(i,j)}_{A_0-a_\alpha,Z_0-z_\alpha}}{\Omega_{A_0,Z_0}}\
\label{eq:yi}
\end{equation}

\noindent
$\alpha=i\,{\rm or}\ j$.

From these probabilities, we may define the correlation between the yields of two selected species as:

\begin{equation}
C_{i,j}\equiv \frac{Y_{i,j}}{Y_i Y_j}=\Omega_{A_0,Z_0}\frac{\Omega^{(i,j)}_{A_0-a_i-a_j,Z_0-z_i-z_j}}{\Omega^{(i,j)}_{A_0-a_i,Z_0-z_i}\Omega^{(i,j)}_{A_0-a_j,Z_0-z_j}}\;.
\label{eq:correlation}
\end{equation}

\noindent
This formula shows that, except for the global factor $\Omega_{A_0,Z_0}$, which is independent of the selected species, this correlation, as defined, does not depend directly on the properties of the species `$i$' and `$j$', which are hold by $\omega_i$ and $\omega_j$.
The latter cancel out in the ratio and, therefore, do not play any role in the expression.
The correlations originate in the characteristics of the partition functions  $\Omega^{(i,j)}_{A_0-a,Z_0-z}$ in the second factor on the right hand side of Eq.\ (\ref{eq:correlation}), whose properties are determined by the composition of the fragmentation modes, after the removal of species `$i$' and `$j$'.

\

\

\end{section}

\begin{figure}[tbh]
\includegraphics[width=8.5cm,angle=0]{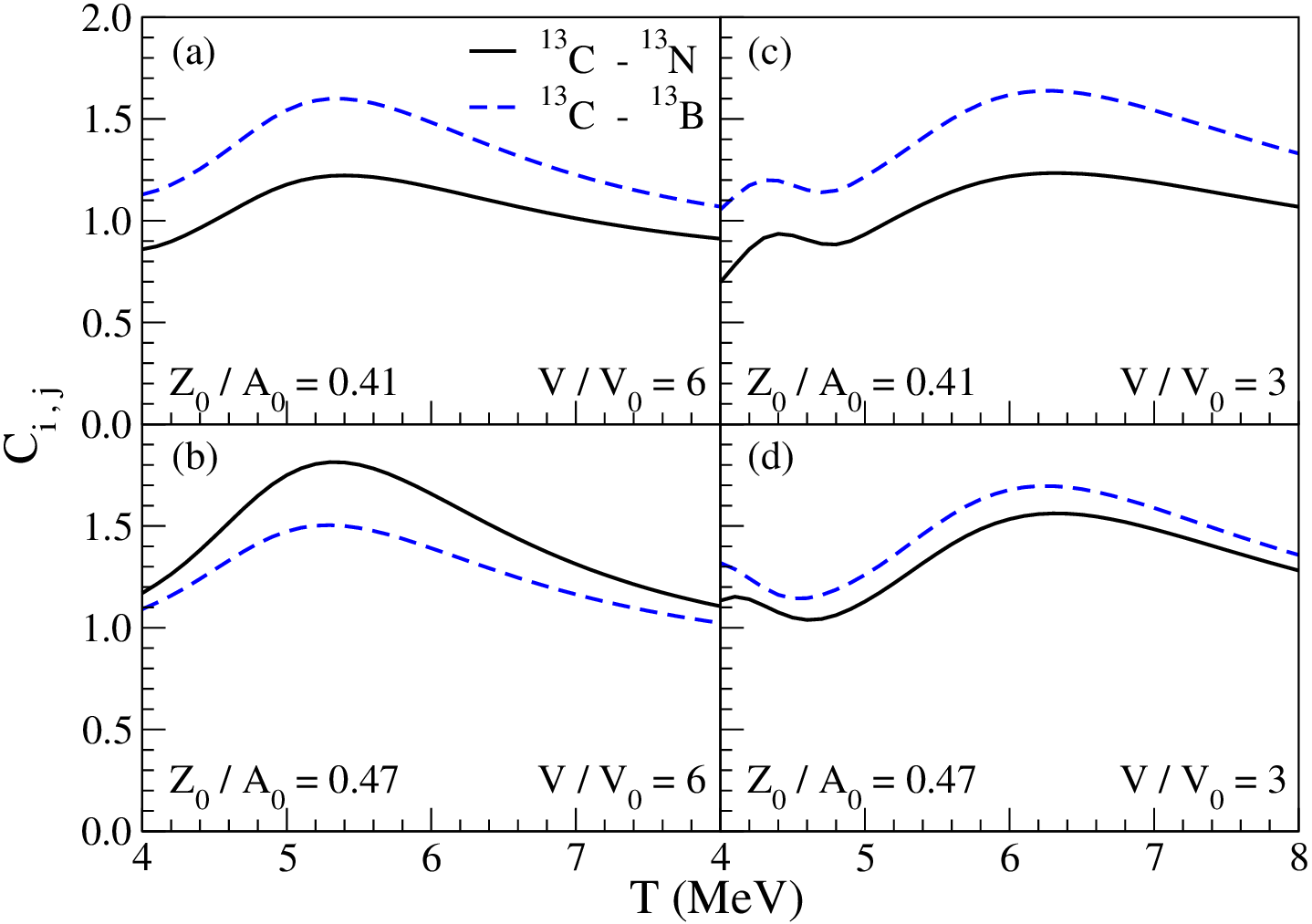}
\caption{\label{fig:cor13v} (Color online) Two-particle correlations as a function of the breakup temperature. In frames (a) and (c) the atomic number of the source is $Z_0=77$ whereas $Z_0=89$ in frames (b) and (d). The breakup volume used in the calculations shown in frames (a) and (b) is $V=6 V_0$ whereas $V=3 V_0$ has been adopted in those exhibited in frames (c) and (d). For details, see the text.}
\end{figure}

\begin{section}{Results}
\label{sect:results}
We apply the correlation defined in the previous section to study the properties of the sources of mass number $A_0=189$ and atomic nubmers $Z_0=77$ and $Z_0=89$.
These sources have been selected because $A_0=189$ corresponds to 80\% of the $^{124}$Xe+$^{112}$Sn system, recently studied in Ref.\ \cite{FragsSizeBorderie2018}.
The different $Z_0$ values are intended to simulate neutron rich and nearly symmetric sources, which allow us to investigate the sensitivity of $C_{i,j}$ to the isotopic composition of the source.
The parameterization for the fragments' binding energies developed in Ref.\ \cite{ISMMlong} is adopted, except where stated otherwise.

Figure \ref{fig:cor13v} shows $C_{i,j}$ for two pairs of fragments: $^{13}$C - $^{13}$N and $^{13}$C - $^{13}$B.
In panels (a) and (b), a breakup volume six times larger than $V_0$ is used.
The isotopic composition of the sources is different in panels (a) and (b).
In the former, a neutron rich source is adopted whereas a nearly symmetric one is assumed in the latter.
One sees that the position of the peak of $C_{i,j}$, which occurs at $T\approx 5.3$ MeV, is fairly insensitive to the species employed and to the isotopic composition of the source.
However, the height of the peak is much higher for the $^{13}$C - $^{13}$B pair than for the mirror nuclei pair $^{13}$C - $^{13}$N if the source is neutron rich.
The opposite happens in the case of the nearly symmetric source.
Since the isotopic composition of the source affects the statistical weight $\Omega^ {i,j}_{A_0-a,Z_0-a}$ only through the species composition within the partitions \footnote{One should recall that, as discussed in sect.\ \ref{sect:model}, the Wigner-Seitz term associated with the homogeneous sphere does not play any role in the canonical formulation.}, this effect is due to the interplay between the symmetry energy of the fragments and the volume terms of the Helmholtz free energy.
This is illustrated in frames (c) and (d) of Fig.\ \ref{fig:cor13v}, which exhibits $C_{i,j}$ obtained with the same parameters as in frames (a) and (b), except for the breakup volume which is $V/V_0=3$, as indicated in the frames.
One sees that, at the smaller breakup volume, the difference between $C_{i,j}$ diminishes if the source's asymmetry is reduced. However, owing to the smaller breakup volume, the changes in the Helmholtz free energy are not large enough to lead to the inversion observed from panels (a) to (b).

On the other hand, it clearly shows that the position of the peak moves to a larger temperature value, being now placed at $T\approx 6.3$ MeV.
Although it is difficult to predict the position at which the peak will occur, our results clearly show that it is fairly sensitive to the breakup volume due to the important funcional dependence of the Helmholtz free energy on this quantity.
Therefore, it may be used to obtain information on the freeze out density.

\

\

\begin{figure}[tbh]
\includegraphics[width=8.5cm,angle=0]{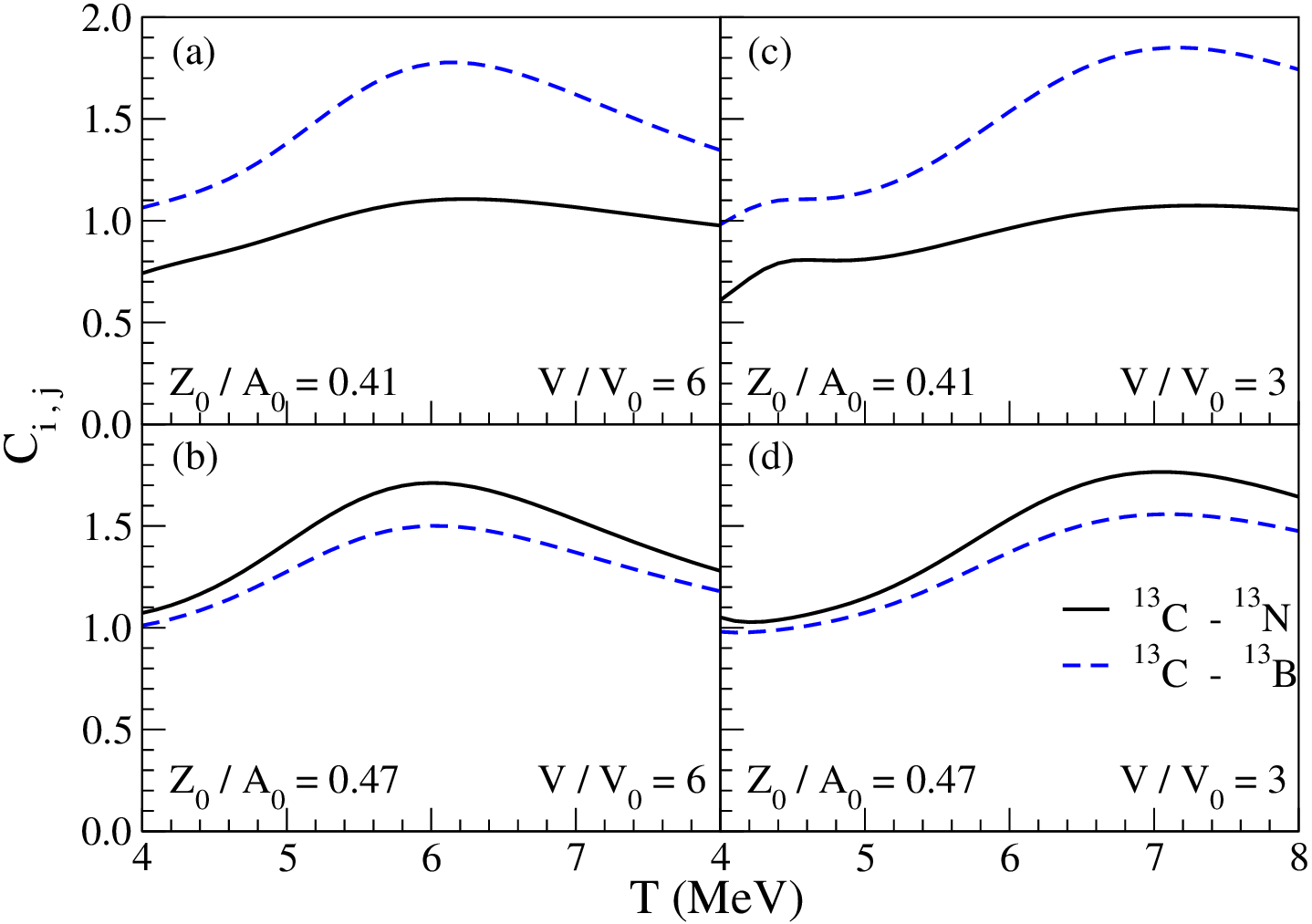}
\caption{\label{fig:cor13vBot} (Color online) Same as Fig.\ \ref{fig:cor13v} for the mass formula used in Ref.\ \cite{Bondorf1995}. For details, see the text.}
\end{figure}

In order to examine the sensitivity of $C_{i,j}$ to the mass formula employed to describe the fragments' binding energies, we have also carried out the calculations with that traditionally employed in the SMM \cite{Bondorf1995}.
The results are displayed in Fig.\ \ref{fig:cor13vBot}, which have been obtained using the same parameters adopted in the previous calculations, except for the mass formula.
The  qualitative features observed previously are also present in Fig.\ \ref{fig:cor13vBot}. 
The noticeable differences are on the quantitative level.
The peaks of $C_{i,j}$ calculated with mirror nuclei are broader, being almost flat for neutron rich sources.
The differences between the results obtained using non-mirror nuclei and mirror nuclei are much larger in the present case.
 The sensitivity to the isotopic composition of the source is also much stronger if one adopts this mass formula.
The position of the peaks are also appreciably affected by the fragments' binding energy.
Now, they are found at $T\approx 6.1$ MeV, for $V/V_0=6$ and at $T\approx 7.1$ MeV for $V/V_0=3$.
These temperatures are about 1 MeV higher than in the previous case.

\

\

\begin{figure}[tbh]
\includegraphics[width=8.5cm,angle=0]{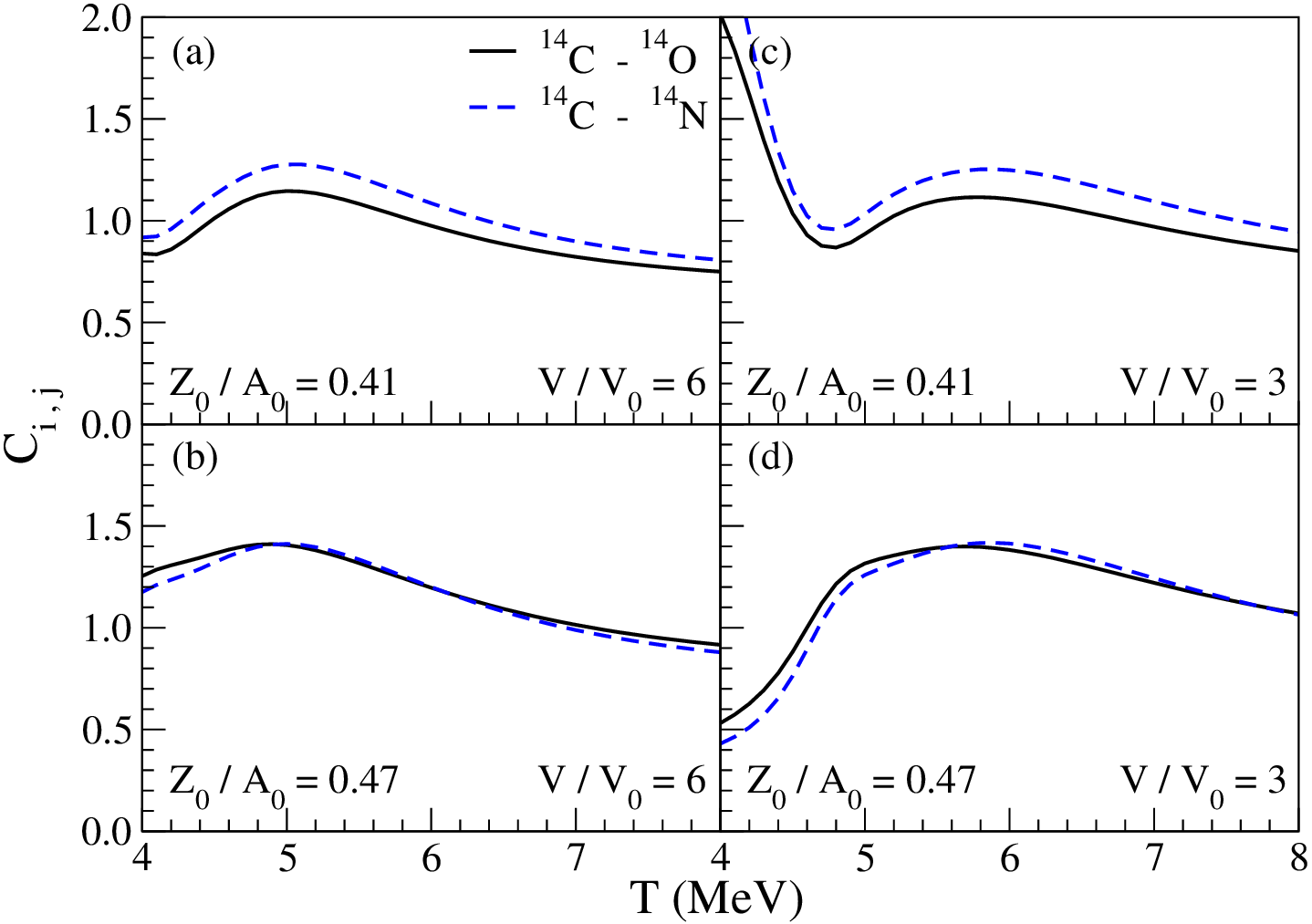}
\caption{\label{fig:cor14v} (Color online) Same as Fig.\ \ref{fig:cor13v} for sources of mass number $A_0=122$ and atomic numbers $Z_0=50$ and $Z_0=77$. For details, see the text.}
\end{figure}

Similar conclusions hold for different isobar pairs and smaller source's size.
To illustrate this point, Fig.\ \ref{fig:cor14v} displays $C_{i,j}$ for the $^{14}$C - $^{14}$O and $^{14}$C - $^{14}$N pairs, produced in the breakup of sources of mass number $A_0=122$ and atomic numbers $Z_0=50$ and $Z_0=77$.
We return to the binding energy prescription developed in Ref.\ \cite{ISMMlong} and adopted in the calculations shown in Fig.\ \ref{fig:cor13v}.
As anticipated, the same trends discussed above are observed in the present case, but the magnitude of the effects are smaller.
This is associated with the smaller sources' volumes.
Furthermore, the peaks occur at $T\approx 5.0$ MeV and at $T\approx 5.8$ MeV for $V/V_0=6$ and $V/V_0=3$, respectively.
This is fairly close to the values obtained with the larger systems and different isobar pairs used in Fig.\ \ref{fig:cor13v}.

Therefore, our results suggest that the two-particle correlations studied in this work may provide valuable information on the breakup density of hot nuclear sources.

\end{section}

\begin{section}{Conclusions}
\label{sect:conclusions}
We have studied the behaviour of a two-particle correlation function, $C_{i,j}$, based on the yields of selecrted isobar pairs produced in the breakup of a hot nuclear source.
Fast recursive formulae to evaluate it have been derived in the framework of the SMM in the spirit proposed in Refs.\ \cite{ChaseMekjian1995,SubalMekjian}.
They have been applied to sources of different isotopic compositions and sizes.
The sensitivity of $C_{i,j}$ to the breakup volume assumed in the statistical calculations has been investigaged within the range bracketed experimentally \cite{PhaseTransitionBorderie2019,freezeOut,freezeOut2,freezeOutPiantelli2005PLB,isoNatowitz2007_2,reviewInterferometryKonrad1990,interferometryBetty1987,interferometryGiuseppe2002,interferometryGiuseppe2003} as a function of the temperature.
The correlations turned out to be appreciably sensitive to the freeze out density and, therefore, our results suggest that they may provide valuable information on this quantity.
Furthermore, as the fragments' energies also affect the properties of $C_{i,j}$, it may be used to obtain information on them as their properties, such as the symmetry energy coefficient, may be affected due to the finite temperature of the system \cite{PhaseTransitionBorderie2019,SymEnergyBaoAnLi2006,isoNatowitz2007_2,symEnergyShatty2007,EOSLi,isoSMMTF,SouliotisCsym2009,TanIsoEOS,TanisoDiff,smmtf1,isotemp}.
We thus suggest that $C_{i,j}$ be used to improve the picture for the freeze out configuration.

\end{section}

\begin{acknowledgments}
This work was supported in part by the Brazilian
agencies Conselho Nacional de Desenvolvimento Cient\'\i ­fico
e Tecnol\'ogico (CNPq), by the Funda\c c\~ao Carlos Chagas Filho de
Amparo \`a  Pesquisa do Estado do Rio de Janeiro (FAPERJ),
a BBP grant from the latter. 
We also thank the Uruguayan agencies
Programa de Desarrollo de las Ciencias B\'asicas (PEDECIBA)
and the Agencia Nacional de Investigaci\'on e Innovaci\'on
(ANII) for partial financial support.
This work has been done as a part of the project INCT-FNA,
Proc. No.464898/2014-5.
We also thank the N\'ucleo Avan\c cado de Computa\c c\~ao de 
Alto Desempenho (NACAD), Instituto Alberto Luiz Coimbra de 
P\'os-Gradua\c c\~ao e Pesquisa em Engenharia (COPPE), 
Universidade Federal do Rio de Janeiro (UFRJ), for the use 
of the supercomputer Lobo Carneiro, as well as the Cloud Veneto, 
where the calculations have been carried out.

\end{acknowledgments}

\bibliography{manuscript}
\bibliographystyle{apsrev4-1}

\end{document}